\documentclass[authoryear,3p,final,11pt]{elsarticle}
\usepackage{amsmath}
\usepackage{amsthm}
\usepackage{xcolor}
\usepackage{amsfonts}

\usepackage{algorithm2e}
\usepackage{listings}

\newcommand{\red}[1]{\textcolor{red}{#1}}

\lstdefinelanguage{pseudo}
{morekeywords={procedure, for, do, if, then, end, Queue, PriorityQueue, function, else, return, to, while},
sensitive=false,
morecomment=[l]{//},
morecomment=[s]{/*}{*/},
morestring=[b]",
}
\theoremstyle{definition}

\newtheorem{definition}{Definition}
\newtheorem{problem}{Problem}
\newtheorem{theorem}{Theorem}

\newcommand{\IGNORE}[1]{}

\journal{the Journal of Pseudosciences}

\begin{document}

\begin{frontmatter}

\title{Detachment Problem -- Application in Prevention of Information Leakage in Stock Markets}

\author[1]{Henri Hansen \corref{cor1}}
\cortext[cor1]{henri.hansen@tuni.fi}
\address[1]{Tampere University}

\author[2]{Juho Kanniainen \corref{cor2}}
\cortext[cor2]{juho.kanniainen@tuni.fi}

\begin{abstract}

In this paper, we introduce Detachment Problem within the context of the well-established independent cascade model. It can be seen as a generalized Vaccination Problem, where we acknowledge that an individual may belong to multiple circles. The aim is to optimally cut the individuals' ties to these circles to minimize the overall information transfer in a social network under a scenario where a limited number of investor-circle pairs can be operated. When an individual is isolated from a particular circle, it leads to the eliminating of the connections to all the members of that circle, yet the connections to other circles remain. 
This approach contrasts with the conventional vaccination problem, which generally involves selecting a subset of vertices for total elimination. In our case, the connections of individuals to the circles are selectively, rather than entirely, eliminated. 

Contextually, this article focuses on private information flows, specifically within networks formed by memberships in circles of insiders. Our quasi-empirical study uses simulated information flows on a truly observable network, and the statistical properties of the simulated information flows are matched with real-world data. In a broader context, this paper presents the Detachment Problem as a versatile approach for optimal social distancing, applicable across various scenarios. 

We propose and define a concept of expected proportional outside influence, or EPOI, as measure of how widespread information leak is. We also implement a greedy algorithm for finding a set of detachments to minimize EPOI. For comparison, we devise a simple heuristic based on minimal cut, to separate the most influential circles from each other. We provide evidence that the greedy algorithm is not optimal, and it is sometimes outperformed by the simple heuristic minimum cut algorithm, However, the greedy algorithm outperforms the cut algorithm in most cases. Further avenues of research are discussed.

\end{abstract}


\end{frontmatter}


\section{Introduction}

The propagation of ideas, fake news, influence, and diseases through social networks has been examined from numerous angles, encompassing fields such as technology adoption, the spread of rumors, viral marketing initiatives, and epidemiology, among others \citep{kempe2003maximizing,nowzari2016analysis,li2018influence,sharma2019combating}. This paper defines and introduces a new problem, which we term the Detachment Problem. 

In the Detachment Problem, an individual is connected to one or more circles. A circle is simply a set which connects all its members in some way. In our data this is the set a board of directors and high level managers of a company, i.e., the set of insiders, but it can be participants in a given meeting, a school class, etc in other applications. The objective is to optimally select a given number of individual-circle pairs and remove the selected individuals from the selected circles, to minimize the information transfer in a social network. When a vertex is isolated from a given circle, it leads to the eliminating its connections to all the members of that circle. However, the connections to other circles remain. We address the detachment problem assuming the dynamics of information spread follow the well-established cascade model, known as the \emph{independent cascade model} \citep{goldenberg2001talk,kempe2003maximizing}.

The Detachment Problem can be seen as a generalized case of the \emph{vaccination} or \emph{immunization problem}, i.e., selecting an individual and either immunize or, in the case of the detachment problem, partially isolate, from a given social network \citep[for vaccination and immunuzation problems, see, for example,][]{pastor2002immunization,gallos2007improving,prieto2021vaccination}. While conventional vaccination approaches focus on immunizing a subset of vertices (individuals), the Detachment Problem differs as it does not entail the complete removal of a vertex. Instead, it involves cutting off its connections to specific circles. Additionally, this problem bears a strong resemblance to the task of identifying influential spreaders or nodes within a network, or influence maximization, as discussed in various studies \citep{kitsak2010identification, chen2012identifying, morone2015influence, lu2016vital}. However, the distinctive aspect of the Detachment Problem lies not in identifying and removing a vertex (individual) but in eliminating its ties with certain circles to which it belongs. 

The framework for the Detachment Problem outlined in the paper is highly generic, with a primary emphasis on methodology. This problem is applicable in a variety of contexts, ranging from the regulation of financial markets to the containment of infectious diseases. To concretize the usefulness of the framework, we focus on private information flows, in particular, in networks that emerge from memberships in circles of insiders. The go-to model we use is the well-known \emph{independent cascade} model (IC), which captures a simplified susceptible-infected-recovered (SIR) model in a network of individuals. The problem of \emph{influence maximization} is that of finding a subset of vertices to "infect" in order to maximize the diffusion of information, and it has received significant attention in the literature. From the perspective of the application, this article addresses the question of preventing the spread of illicit information,
which might lead to, for example, insider trading.

We study the problem by modeling a particular network of insiders, i.e., board members and key executives of companies listed in the Helsinki Stock exchange in the year 2009. These persons form a social network, and their known connections come from being members of the same inner circle in a company. As many key personnel hold key positions in multiple companies, this network represents a plausible medium through which inside information might spread \citep{ahern2017information,berkman2020inside,baltakiene2022trade,baltakys2023predicting}. Publicly traded companies are obligated to announce important information publicly, but the insiders of these companies gain access to such information before the information is made public. The use of such information to make profitable trades 
in the stock market before the information is publicized is known as \emph{insider trading}, and it is illegal in most jurisdictions. Insider trading is estimated by various sources to result in massive illicit gains and as such create a pernicious and hard-to-detect cost in the economy, which makes our research question a pertinent one. What is more, our model and results point to regulatory recommendations and allow us to quantify the effect of the implementation of said recommendations.

\section{Model and Problem formulation}

\subsection{Induced network}
\begin{definition}
Let $V$ be a set of vertices, and $\mathcal I = \{I_1, I_2, \ldots I_k\}$ where
$I_i \subseteq V$ be a collection of sets of vertices called \emph{circles}.
The \emph{network induced by} $\mathcal I$ is the graph $(V,E)$ where
$$V = \bigcup_{I_i \in \mathcal I} I_i,$$
$$E = \{ (u,v) \mid \exists i: u \in I_i \wedge v \in I_i\}.$$
\end{definition}
In other words, there is an edge between two vertices if there is a circle in which both belong. 

We associate a weight $w_{u,v} \in [0,1]$ for each $(u,v) \in E$, to denote a probability that $u$ passes information forward to $v$. Then we call
$(V,E,w)$ an \emph{information flow network}.

\subsection{The Independent Cascade Model}

The so-called Independent cascade model~\citep{goldenberg2001talk,kempe2003maximizing} describes a simple process by which information starting from an initial set of
vertices spreads through such a network. It is a variant of the {\bf S}usceptible - {\bf I}nfected - {\bf R}ecovered, or SIR-model, in which each vertex is infected at most once in some time step, will attempt to infect all of its neighbours in the next, and will be recovered and totally inert in all the subsequent time steps.   

Let $A_0 \subseteq V$. We define a \emph{cascade from $A_0$}  as random sequence 
$\gamma(A_0) = \langle A_0, A_1, \ldots, A_n \rangle$ where each $A_i \subseteq V$ and for each $i = 1, \ldots n$ we have
$$A_i \cap \displaystyle \bigcup_{j < i} A_j = \emptyset$$
and for every $v \in A_{i+1}$ there is some $u \in A_i$ such that 
$(u,v) \in E$ and $w_{u,v} \geq 0$. 
The cascade is generated by the following process, which we call the 
\emph{independent cascade process} or IC-process:
\begin{enumerate}
    \item The initially infected set is the set $A_0$, and set $i = 0$.
    \item If $A_i$ is empty, then stop; the cascade is finished and $n = i-1$.
    \item Let $A_{i+1}$ be a set. 
    \item For each $u \in A_i$ and $v$ such that $(u,v) \in E$ and $v$ is
    not yet infected: Add $v$ to the set $A_{i+1}$ with probability $w_{u,v}$;
    if $v$ is added to $A_{i+1}$ mark it as influenced. 
    \item Set $i = i+1$ and go to 2. 
\end{enumerate}
Given an information flow network and an initial set $A_0$, we denote by 
$$\sigma(A_0) = E\left(\left|\bigcup_{A \in \gamma(A_0)} A\right|\right).$$ 
I.e, $\sigma(A_0)$ is the expected number of vertices in
the cascade and the function $\sigma$ is known as the \emph{influence function}. 
There is no closed-form solution to the computation of the influence function (that
we know of), and it has been proposed that the computation of the influence function is P\#-complete(\cite{shakarian2015independent})

\subsection{The Detachment problem}

Let $\mathcal I = \{I_1, I_2, \ldots I_k\}$ be a collection of circles and let $G = (V,E,w)$ be the information
flow network induced by $\mathcal I$. Let $p$ be a distribution over $\mathcal I$,
or, equivalently, a distribution over the set $\{1, \ldots, k\}$. 
We define the \emph{expected proportional outside influence} or EPOI of the information flow network as 
$$
\rho_\mathcal{I} = \sum_{I \in \mathcal I} p(I)\cdot \left(\frac{\sigma(I) - |I|}{|V \setminus I|}\right)
$$
In other words, the EPOI of the network is the expected proportion of non-circle members influenced when a random circle is the source of a cascade. 

Given a collection of circles $\mathcal I = \{I_1, \ldots,I_k\}$,
we define the \emph{detachment} of a vertex $v$ from a given circle $I_i$ such that
$v \in I_i$, as the collection 
$$\det(\mathcal{I},(v,I_i)) = 
\{I_1, \ldots, I_{i-1}, I_{i}\setminus\{v\}, I_{i+1}, \ldots, I_k\}$$
and we generalize this to detaching a 
set of pairs $S = \{(v_1, I_{i_1}), \ldots, (v_m, I_{i_m})\}$ in 
the obvious way, i.e, 
$$\det(\mathcal{I}, S) = 
\det(\det(\cdots \det(\mathcal{I}, (v_1,I_{i_1})) \cdots, (v_{m-1}, I_{i_{m-1}}) ),  (v_m, I_{i_m}) )$$

We now define the \emph{$m$-detachment problem}:
\begin{problem}\label{p:det}
Find the 
detachment $S$ such that $|S| = m$ and
$\rho_{\det(\mathcal{I}, S)}$ is minimized. 
\end{problem}
\section{Algorithms and Transformations}

\subsection{Bridge-block graph}

\begin{definition}
    Let $V$ be a set of vertices and let $\mathcal I$ be a collection of circles. 
    The \emph{Bridge-Block network}  (BBN) induced by $\mathcal I$ is a bipartite graph
    $(\mathcal I, U, F)$, where
    \begin{enumerate}
        \item $\mathcal I = \{I_1, \ldots, I_k\}$, is the set of \emph{blocks},
        i.e., each circle is called a block of the BBN.
        \item $U = \{v \in V \mid \exists i, j: i \neq j \wedge v \in I_i \cap I_j \}$,
        is the set \emph{bridges}, i.e., each vertex that belongs to more than one circle,
        is called a bridge of the BBN
        \item $F = \{(I, v) \mid I \in \mathcal I \wedge v \in U \wedge v \in I\}$, i.e., 
        a block and a bridge have an edge between them, if and only if the bridge belongs to 
        the block.
    \end{enumerate}
\end{definition}
We will use the bridge-block graph in our algorithms, and make use of the fact that a detachment of $(v,i)$ from $\mathcal I$, is equivalent to removing exactly one edge, namely $(I_i, v)$ from the bridge-block graph induced by $\mathcal I$.

\subsection{Greedy algorithm}
\begin{algorithm} 
\KwData{Circles $\mathcal I$, edge weights $w$, initial distribution $p$, number of detachments
$m$}
$\rho_{min} \gets \infty$\;
\For{$i = 1$ to $m$}
{
    $(\mathcal I, U, F) \gets BBN(\mathcal I)$\;
    \For{ each $(I,v) \in U$}
    {
        $\mathcal I' = \det(\mathcal{I}, (v,I))$\;
        \If{$\rho_{\mathcal I'} < \rho_{min}$}
            {
                $\rho_{min} \gets \rho_{\mathcal I'}$\;
                $\mathcal I^* \gets \mathcal I'$\;
            }
    }
    $\mathcal{I} \gets \mathcal I^*$\;
}
\KwResult{The network $\mathcal I^*$ with EPOI $\rho_{min}$}
\vspace{0.4cm}
\caption{Greedy algorithm} \label{alg:greedy}
\end{algorithm}

A simple greedy algorithm for solving Problem~\ref{p:det} is given in Algorithm~\ref{alg:greedy}. 
The strategy is very simple, but we present it here for illustrative purpose. 
The algorithm itself is amenable to many heuristics and optimizations, such as omitting vertices that have a particularly low centrality measure. We will discuss these heurisics in a further chapter. 

\subsection{Min-cut}
Because the computation of the EPOI is time consuming with Monte Carlo, there is definitely a rationale for exploring methods that do not need the evaluation of the EPOI, and rely purely on heuristics of the graphs instead. 

\begin{algorithm}
\KwData{Circles $\mathcal I$, weights $w$ and initial distribution $p$}
Choose $\mathcal I_1 \subseteq \mathcal I$ and $\mathcal I_2 \subseteq \mathcal I$ such that $\mathcal I_1 \cap \mathcal I_2 = \emptyset$ and $\sigma(\mathcal I_1)$ and
$\sigma(\mathcal I_2)$ are large\;
Heuristically determine weights for the BBN\;
Add vertex $s$ and edge $(s, \mathcal I_1)$, with a large weight\;
Add vertex $t$ and edge $(\mathcal I_2, t)$ with a large weight \;
Find \textsc{Min-Cut} for the BBN that separates $s$ and $t$
\KwResult{The network $\mathcal I'$ that has 2 components}
\caption{Minimum Cut}\label{alg:cut}
\end{algorithm}

\begin{theorem}
A (minimum) cut that separates two circles in a BBN is equivalent to a detachment\;
\end{theorem}
The theorem is obvious, because in a BBN, all paths alternate between a block and a bridge. 
The upside of Algorithm~\ref{alg:cut} compared with Algorithm~\ref{alg:greedy} is that it does not require the calculation of the EPOI at each step. 

\section{Empirical and random data}

\subsection{Empirical data}
We will make use of data that comes from insider registries of companies publicly listed  on the Helsinki Stock exchange. 
We use the year 2007 as our benchmark year, i.e., $I_i$ is chosen as
the set of insiders for the company $i$ on the last day of 2006. 
$I_i$ is the set of people who are considered insiders of a given company $i$.   
The BBN of the insiders is not fully connected, so we only choose the companies that are in the largest connected component of the generated BBN. 
Our data has $|\mathcal I|=106$ companies, with a total of $|V| = 967$ insiders, of whom
$|U| = 140$ are bridges, i.e., individuals who are insiders in more than one company. 

\subsection{Random data generation}

The process of generating randomized input for the algorithm involves a systematic procedure designed to mimic the statistical properties of the empirical data and with some modifications as to the parameters.  We fixed the ratios 
$$\alpha = \frac{|V|}{|\mathcal I|},  ~
\beta =  \frac{|V|}{|U|}, ~
\gamma = \frac{|V|}{|F|}$$
The generative process generates the circles via a distribution that is a mix of uniform distibution and preferential selection, using a coefficient  
$0 \leq \rho \leq 1$.
The circle assignment was a four-stage process. 
\begin{enumerate}
    \item 
The first process assigns 
each element a unique circle as follows:
\begin{enumerate}
    \item First we assign one element to each of the circles.
    \item Let $A^(k)$ be the set of assigned elements on iteration $k$. 
    One unassigned element is picked, and the circle for it is chosen so that
    the probability of being assigned into circle $I^{(k+1)}_i$ is 
    $$(1-\rho) \frac{1}{|\mathcal I|} + \rho \frac{I^{(k)}_i}{A^{(k)}}$$
    \item Repeat until each element has been assigned a circle. 
\end{enumerate}
\item The BBN is then forced into being connected by generating $|\mathcal I| - 1$ bridges
\begin{enumerate}
    \item An element is picked using uniform distribution, and its circle is
    chosen as the root circle. Note that this means larger circles have higher probability of being chosen. 
    \item At every iteration, a random element $u$ that is not yet connected, is chosen and added to a randomly chosen connected circle, thereby creating a bridge and
    connecting all the elements from its the circle to the graph. 
    \item this is repeated until all circles are connected to the root. 
\end{enumerate}
\item The rest of the bridges are then created by choosing non-bridge elements 
at random and adding them to random circles. 
\item The bridges are then added to further circles by choosing 
circles using the same mixed distribution as in the first stage.  
\end{enumerate} 
\begin{figure}
    \centering
    \includegraphics[width=\textwidth]{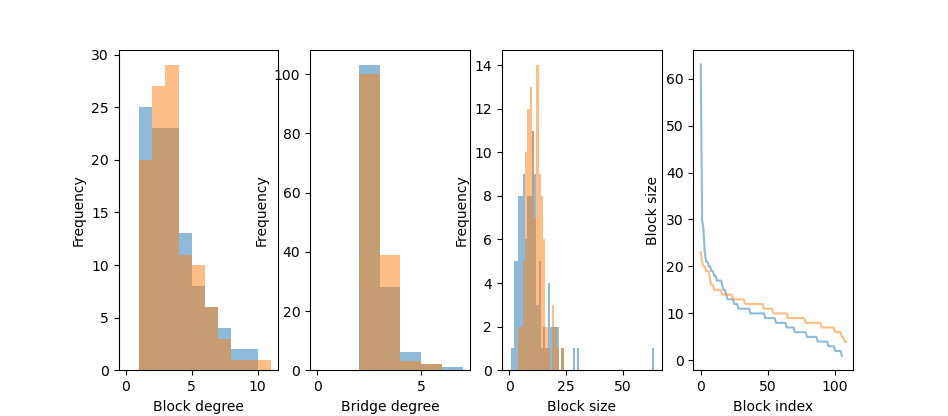}
    \caption{Comparison of empirical (blue) and generated (brown) graph statistics}
    \label{fig:comparison}
\end{figure}
Figure~\ref{fig:comparison} shows graphically a comparison of some graph characteristics of empirical and generated graphs, with $\alpha, \beta$, and $\gamma$ chosen to be equal to those measured empirically, and  $\rho = 0.3$. 
The empirical data had three abnormally large circles that this generative model fails to march, but otherwise the properties are very similar. 

The transmission probabilities between nodes were generated from a Beta-distribution 
with parameters $\alpha = 20, \beta = 80$, which gives the expected value of $w$ at $\mu = 0.2$, and a standard deviation $\sigma \approx 0.04$. We empirically determined that the expected EPOI is very robustly about 0.79 over all the random graphs in our class, including the empirical data.

\section{Results}

We generated 96 random graphs with number of vertices ranging from 100 to 1980  with the process described in the previous section. 
The objective was to gauge the robustness and scalability of our two methods. 
We observed a relatively stable EPOI over our data, and a modest but stable deterioration of the algorithms as the size of the graph increases, which is to be expected.  Below we report the result for a sample of 15 of these
random graphs. 

\begin{table}
\centering
\caption{Numerical results}
\label{tab:num_results}
\begin{tabular}{lrrrrr}
\hline
{n} &  Min-cut &  EPOI cut &  EPOI greedy &  std of EPOI &  base EPOI \\
\hline
280  &             2 &     0.766 &        0.705 &          0.000 &      0.809 \\
320  &             3 &     0.720 &        0.641 &          0.000 &      0.789 \\
680  &             1 &     0.768 &        0.720 &          0.000 &      0.783 \\
740  &             2 &     0.756 &        0.713 &          0.000 &      0.770 \\
860  &             2 &     0.779 &        0.736 &          0.000 &      0.790 \\
880  &             1 &     0.766 &        0.754 &          0.000 &      0.779 \\
900  &             2 &     0.785 &        0.755 &          0.000 &      0.798 \\
1000 &             1 &     0.789 &        0.783 &          0.000 &      0.800 \\
\red{1020} &             \red{3} &     \red{0.734} &       \red{0.742} &          \red{0.000} &      \red{0.786} \\
1040 &             1 &     0.766 &        0.743 &          0.000 &      0.775 \\
1500 &             1 &     0.782 &        0.775 &          0.000 &      0.790 \\
1860 &             2 &     0.777 &        0.757 &          0.000 &      0.783 \\
1920 &             3 &     0.766 &        0.744 &          0.000 &      0.778 \\
1960 &             3 &     0.754 &        0.736 &          0.000 &      0.774 \\
1980 &             1 &     0.789 &        0.783 &          0.000 &      0.793 \\
\hline
Empirical (967)     &          3 & 0.712 & 0.650 & 0.000 & 0.729 \\
\hline
\end{tabular}
\end{table}

The greedy algorithm performed better than the cutting algorithm in all but one case. In order to better understand what created the difference, we conducted a number of comparisons of graph characteristics but found no explaining factor.
Further research is warranted in this respect. 

\section{Discussion}

This short paper demonstrates the idea of the detachment problem, and is still under work. We try to present here the main ideas of detachment, the sort of networks that it might work on, and one heuristic approach. 

Full combinatorial optimization is still intractable in this case, and the cutting algorithm does not seem to provide a satisfactory heuristic even compared with the greedy algorithm. Our tentative results already demonstrate that in realistic problems, the greedy algorithm is not optimal, which is not surprising. What may be surprising, is that a very simple heuristic algorithm outperforms it on a problem with realistic characteristics.

\bibliography{mybibfile}

\end{document}